\documentstyle[camera,prbplug,psfig]{article}
\begin{document}

\title{Energy Spectrum of Spin Fluctuations in Superconducting
La$_{2-x}$Sr$_x$CuO$_4$ ($0.10 \leq x \leq 0.25$)}

\author{Chul-Ho Lee\thanks{{\it Permanent address:} Electrotechnical Laboratory,
Umezono 1-1-4
Tsukuba, 305-8568, Japan}, Kazuyoshi Yamada\thanks{{\it Permanent address:} Institute for Chemical
Research, Kyoto University, Uji 611-0011, Japan}, Yasuo Endoh \\
{\it Department of Physics, Tohoku University, Aramaki Aoba, Sendai
980-77, Japan} \\
Gen Shirane\\{\it Department of Physics, Brookhaven National
Laboratory, Upton, NY 11973-5000, USA}\\
R.\@ J.\@ Birgeneau, M.\@ A.\@ Kastner, M.\@ Greven\thanks{{\it Permanent address:} Dept. of Applied Physics,
Stanford University,
Stanford, CA 94305, USA}, Y-J.\@ Kim\\
{\it Department of Physics and Center for Materials Science and
Enginnering, Massachusetts
Institute of Technology, Cambridge, MA 02139, USA}}
\date{November 12, 1999}
\maketitle

\begin{abstract}
The energy spectrum of incommensurate spin
fluctuations in superconducting Sr-doped La$_2$CuO$_4$ has been
studied by inelastic neutron scattering experiments.  An energy gap in
the spin excitation spectrum is observed in the superconducting state
of optimally doped (x=0.15) and slightly overdoped (x=0.18)
samples.  At temperatures well below T$_{c}$, the incommensurate peaks
diminish rapidly in intensity with decreasing energy below $\sim$8meV and
merge into the background below $\sim$3.5meV for x=0.15 and $\sim$4.5meV
for x=0.18.  For
both samples, the energy spectrum of the  q-integrated spin susceptibility,
$\chi^{\prime
\prime}(\omega)$, exhibits an enhancement around 7meV, which is caused by a
broadening
in the momentum width of the incommensurate peaks.  The gap-like structure in
the energy spectrum and the enhancement of $\chi^{\prime \prime}(\omega)$
survive at T$_{c}$.  On the other hand, for both underdoped
(x=0.10) and heavily overdoped
(x=0.25) samples there is neither a clear energy gap nor an
enhancement of $\chi^{\prime \prime} (\omega)$ below T$_{c}$.\\
\\KEYWORDS: La$_{2-x}$Sr$_x$CuO$_4$, high-T$_c$ superconductor, spin fluctuation,
neutron scattering
\end{abstract}

\section{Introduction}
\label{sec:level1}
The interplay between magnetism and superconductivity is one of the most
interesting and important issues in the physics of the
cuprate high temperature superconductors [1].  For more than ten years,
experimental studies on the spin fluctuations in these materials have
been performed by neutron scattering and NMR measurements.  NMR first
revealed a suppression of the low-energy spin excitations below what is
called the spin gap temperature [2].  In the underdoped region, it is
thought that above T$_c$ a pseudo-gap opens up in the spin fluctuation
spectrum.  Since the spin-gap state is believed to be related to the
pairing mechanism, a large number of experimental and theoretical studies
have focused on the origin of the spin gap.

Unlike NMR, neutron scattering can determine both the momentum and the
energy dependence of the magnetic excitations, providing information
about the anisotropy of the energy gap of the high-T$_c$ cuprates.  In
fact, neutron scattering has been used to observe the spin-gap of
YBa$_2$Cu$_3$O$_{7-y}$ by determining the energy spectrum of the
dynamical spin susceptibility $\chi^{\prime \prime}(q,\omega)$ around $(\pi,
\pi)$ as a function of T [3,4,5].  However, the energy spectrum of the spin
fluctuations and its doping dependence in La$_{2-x}$Sr$_x$CuO$_4$ have
not yet been determined over a wide range of x, most especially for x
above the optimal doping value.  This is in part because it is difficult
to separate the weak magnetic signals from the non-magnetic background
originating from phonons for example.  Further, it is difficult to grow
large homogeneous single crystals over a wide doping region.

Several years ago an overview of the low energy spectrum
of $\chi^{\prime \prime}(q, \omega)$ associated with the incommensurate peaks of
La$_{1.85}$Sr$_{0.15}$CuO$_4$ was presented [6].  Shortly thereafter, a
well-defined energy gap of $\stackrel{>}{\sim}$ 3.5meV was observed in the
superconducting state of the optimally Sr-doped La$_2$CuO$_4$ [7].  The
latter measurements were limited to low energies
$\stackrel{<}{\sim}$ 6meV.  For higher energies, pulsed neutron
scattering measurements were performed on La$_{1.85}$Sr$_{0.15}$CuO$_4$
and revealed a broad peak between
$\omega = 40$ and 70meV [8] in $\chi^{\prime \prime} (\omega)$.  The dynamical
spin susceptibility was integrated over q around the incommensurate peaks.
This result was recently reconfirmed by more comprehensive experiments
and analysis on the same crystals [9].

However, Hayden {\it et al.} [10] reported a somewhat sharper peak in the
energy spectrum of $\chi^{\prime \prime} (\omega)$ at around $\omega =
20$meV in
La$_{1.86}$Sr$_{0.14}$CuO$_4$ by combining data taken by pulsed
neutron scattering with that of 3-axis neutron spectroscopy.  
Furthermore, the latter group studied the energy and temperature dependence
of $\chi^{\prime \prime} (\omega)$ using the same crystal and suggested
that they
were observing the effects of a nearby quantum critical point at T=0 K.  They
reported that the q-width of the spin fluctuations monotonically increased with
increases in either temperature or energy [11] although their observed
correlation length at low temperatures was in fact quite short in apparent
contradiction with the quantum critical interpretation.  They found no evidence
for a gap in the spin fluctuation spectrum above T$_c$, in conflict
with NMR measurements [12].

To reconcile these apparent inconsistencies a more systematic neutron
scattering study is required.  In particular, it is important to
measure the doping dependence of the energy spectrum over a wide
doping region.  To accomplish this we have grown large single crystals
of superconducting Sr-doped La$_2$CuO$_4$ over a wide range of Sr
concentrations extending from the underdoped to the highly overdoped
region.  Although the energy region  in the present study is still
limited to that of conventional low energy 3-axis neutron scattering
measurements, the doping dependence provides new insights into the
spectrum of the spin fluctuations in this system.  We describe the
experimental details in section \ref{sec:level2}, the results and data analysis are
presented in section \ref{sec:level3}, a discussion is given in section \ref{sec:level4}
and a brief summary is presented in section \ref{sec:level5}.

\section{Experimental Detail}
\label{sec:level2}
Single crystals of La$_{2-x}$Sr$_x$CuO$_4$ have been grown by the
traveling solvent floating zone (TSFZ) method using lamp-image
furnaces [13,14].  The furnaces have been improved in order to tune the
temperature gradient near the molten zone.  We have found that a steep
temperature gradient around the liquid allows us to grow homogeneous
large single crystals.  As-grown single crystals have been kept under
oxygen flow at $900^{\circ}$C for 50 hours to remove oxygen defects.

To characterize the Sr content and homogeneity we determined the
phase transition temperature (T$_s$) between the high temperature
tetragonal (HTT) and the low temperature orthorhombic (LTO) phases by
neutron diffraction.  The values for T$_s$ of our single crystals
[15] agree quite well with those previously deduced from powder
data [16,17,18]. T$_s$ depends primarily on the Sr concentration
and specifically a small amount of oxygen non-stoichiometry does not
change T$_s$ appreciably [19,20].  Therefore, the agreement of the measured
T$_s$ for our single crystals with the results obtained from measurements on
powders confirms the Sr concentration of our single crystals. The values of the
Sr concentration measured by electron probe microanalysis were also
consistent with the results from the T$_s$ measurements within the
instrumental resolution.

Superconducting magnetic shielding effects were measured by a SQUID
magnetometer in a magnetic field of 10 $Oe$.  The onset temperature of
the superconducting transition (T$_c$) in the $x=0.15$ single crystal,
T$_c$=37.5K, is almost the same as the maximum value found in powder
samples [18,21].  For $x=0.10, 0.18$ and 0.25 the values of T$_c$ =
29K, 36.5K and 15K, respectively, also agree with those of powder
samples [15].  Since T$_c$ is very sensitive to any
deviation from stoichiometry in the oxygen concentration [20,22], this
agreement
suggests that the oxygen content and ordering of our crystals is nearly
optimal.

Normal state magnetic susceptibilities were also measured using a
SQUID magnetometer with a magnetic field of 1T.  We observed a broad
peak at a temperature T$_{max} $=200K for x=0.18 which depends on both
the Sr and the oxygen concentration [22]. T$_{max}$ of the single
crystal is in good agreement with that of a powder sample with the same
Sr concentration [22, 23, 24].  Lattice parameters of pulverized single
crystals measured by x-ray diffraction at room temperature are also
consistent with previous data [15].

Inelastic neutron scattering measurements were performed with the
triple-axis spectrometer, TOPAN at JRR-3M in the Tokai Establishment
of JAERI for crystals with x=0.15, 0.18, and 0.25 and with H7 at
the HFBR at Brookhaven National Laboratory for crystals with $x=0.10$
and 0.15.  The incident (final) neutron energy was typically fixed at
E$_i$ (E$_f$) = 14.75meV for TOPAN and E$_i$ (E$_f$) = 14.7meV for
H7.  A typical sequence of the horizontal collimators was
40'-100'-S-60'-B for TOPAN and 40'-40'-S-80'-80' for H7 where S
denotes the sample position.  Pyrolytic graphite crystals were used
both as monochromator and analyzer.  A pyrolytic graphite filter was
used to reduce the intensity of higher order neutrons.  Moreover, in
TOPAN, a sapphire crystal was inserted to reduce significantly the flux of high
energy neutrons.  The single crystals were mounted in an Al container filled
with He gas as a heat exchanger.  For cooling, a closed cycle He
refrigerator was used.

\section{Results}
\label{sec:level3}
The four panels in Fig. 1 show representative q-spectra of the magnetic
fluctuations measured at and below T$_c$.  Constant-energy scans were
performed through the two peaks at ($\pi,\pi(1 \pm \delta)$) as
illustrated in the inset of panel (c).  For $x=0.18$, due to some
\begin{figure}
	  \centerline{\psfig{file=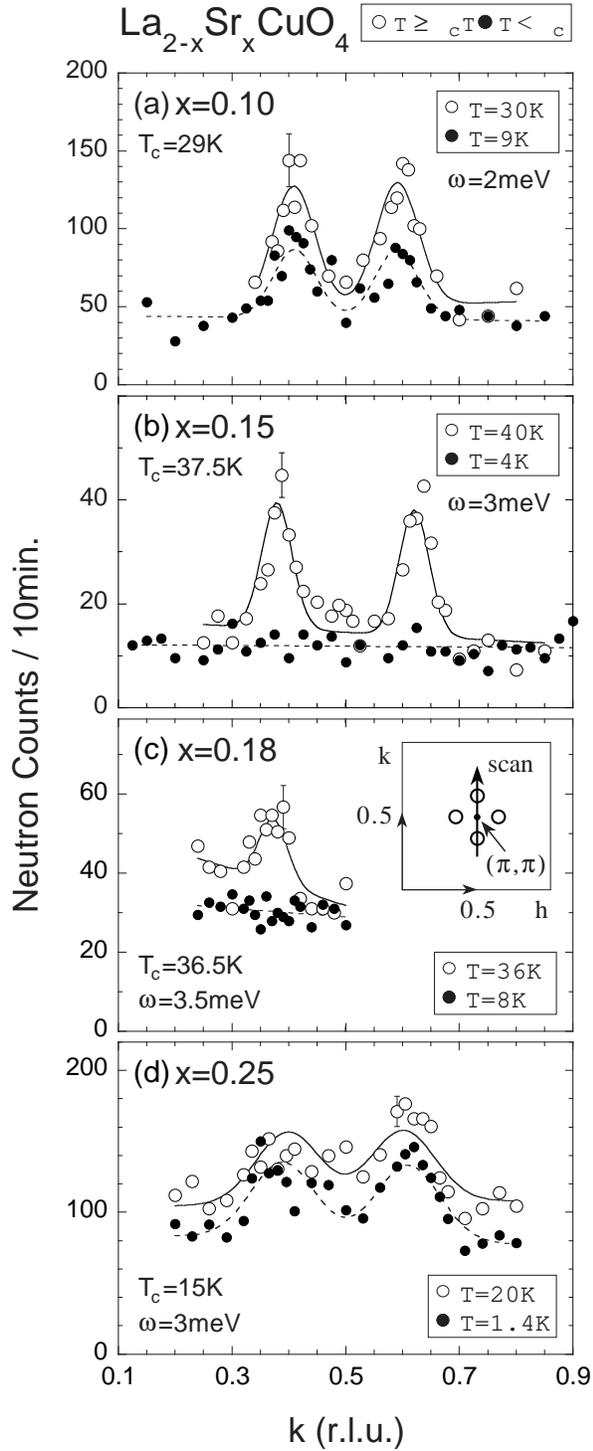,width=\columnwidth}}
	  \caption{Constant energy-scan across ($\pi,\pi$) as indicated in the
inset of panel (c) for samples with (a) x=0.10 (b) x=0.15 (c)
x=0.18 and (d) x=0.25 below and at T$_c$.  All solid lines and dashed
lines for $x=0.10$ and 0.25 are the results of least square fits to the
data points using two Lorentzian functions convoluted with the
instrumental resolution.}
	  \label{f.RvsT-Lt}
	\end{figure}
non-magnetic contamination in the scan, we present only one of the
peaks, at ($\pi,\pi(1-\delta)$).  For  T $\stackrel{>}{\sim}$ T$_c$
well-defined peaks are observed at incommensurate positions for all
samples. Below T$_c$, the peaks at 3 meV and 3.5 meV vanish into the background
for $x=0.15$ and 0.18 respectively.  As is shown by the energy spectra
discussed below, this dramatic change occurs due to the opening of an
energy gap
in the spin excitation spectrum at the same temperature as that for
the onset of superconductivity.  In contrast, for $x=0.10$ and 0.25
substantial peak intensities at low energies are observed even at temperatures
well below T$_c$

We present the results of the same q-scan as that shown in Fig. 1 at several
different energies for x=0.10 and 0.15 in Figs. 2 and 3
\begin{figure}
	  \centerline{\psfig{file=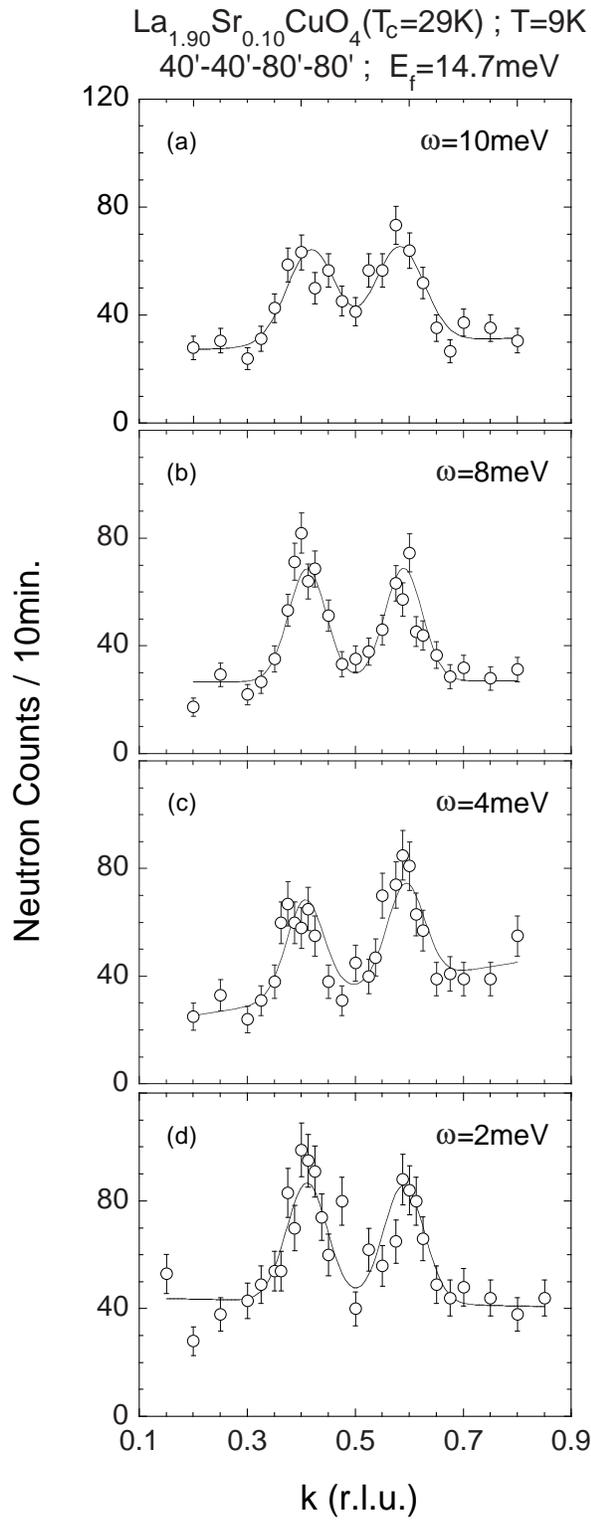,width=\columnwidth}}
	  \caption{Q-spectra at different energies obtained by
	  the same scan as in
Fig. 1 for x=0.10 below T$_c$}
	  \label{f.RvsT-Lt}
	\end{figure}
respectively.  Below T$_c$, the peak intensities for x=0.10 are
comparable at all energies between 2 and 10 meV.  On the other hand,
for x=0.15 the intensities dramatically decrease with decreasing
energy and no peak remains outside of the errors at 3 meV.  Although our
collaboration has reported the observation of an energy gap for
\begin{figure}
	  \centerline{\psfig{file=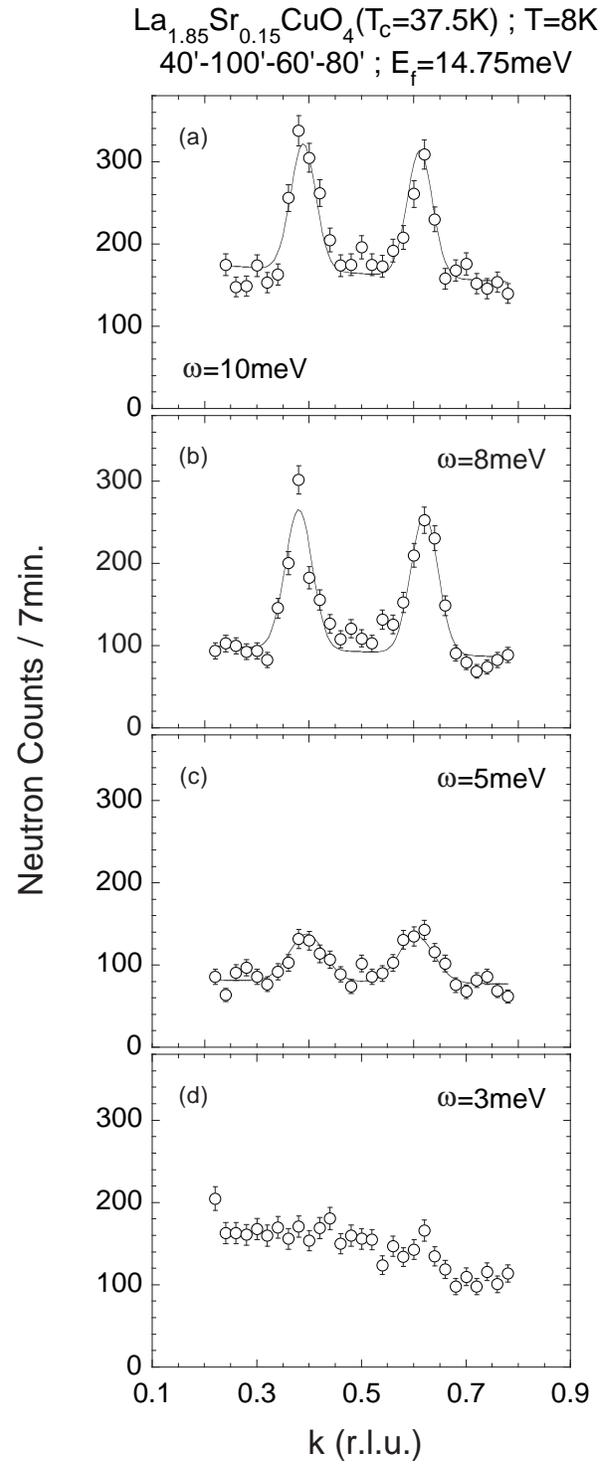,width=\columnwidth}}
	  \caption{Q-spectra at different energies obtained by the same scan as in
Fig. 1 for x=0.15 below T$_c$.}
	  \label{f.RvsT-Lt}
	\end{figure}
x=0.15 previously [7] the change in peak intensities below $\sim$7 meV
appears to be more dramatic in these experiments.  A similar energy
dependence of the incommensurate peak intensity is found for x=0.18
as shown in Fig. 4(c).

The q-spectrum of the incommensurate peaks was
fitted using the following dynamical structure factor S(q,$\omega$)
convoluted with the instrumental resolution function:
\begin{equation}
S(q,\omega)=\frac{I}{1-exp(- \frac{\omega}{\kappa_BT})} \cdot \chi^{\prime
\prime}(q, \omega)
\end{equation}
\noindent with

\begin{equation}
\chi^{\prime\prime}(q, \omega) = A_{\omega} \sum_{\delta=1,4} \left\{
\frac{\kappa_{\omega}}{|q-q_{\delta}|^2+\kappa_{\omega}^2} \right\}
\end{equation}

\noindent where $q_g, \kappa_{\omega}, k_B$  and $A_{\omega}$ are the peak
positions of the magnetic peaks around $(\pi, \pi)$, the q-width at a
given $\omega$, the Boltzman constant and an overall scale factor,
respectively.  $\kappa_{\omega}$ is assumed to be isotropic in the Cu$O_2$
plane.

The imaginary part of the q-integrated dynamical susceptibility
averaged over the Brillouin zone is defined as:
\begin{equation}
\chi^{\prime \prime} (\omega) / \omega=\int \chi^{\prime
\prime}(q,\omega)dq / (\omega \cdot \int dq)
\end{equation}

\noindent Note that in this case, the q-integration is carried out for the four
incommensurate peaks around $(\pi, \pi)$ over the Brillouin zone.

In Fig. 4, we show the energy-dependence of the peak intensities at the
incommensurate positions for $x=0.10, 0.15, 0.18$ and 0.25 together
with the resolution-corrected
$\chi^{\prime \prime}(\omega)/\omega$, which is obtained by fitting the
q-spectrum
\begin{figure}
	  \centerline{\psfig{file=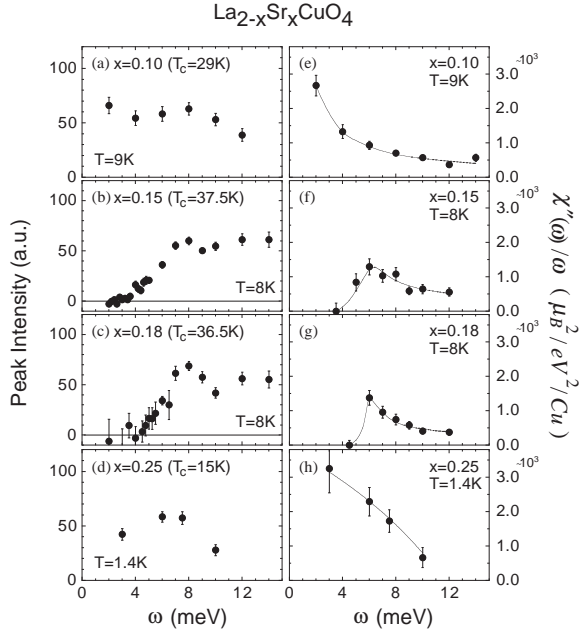,width=\columnwidth}}
	  \caption{Energy dependence of the intensity at the incommensurate peak
position below T$_c$ for $x$=0.10(a), 0.15(b), 0.18(c) and 0.25(d).
Panels (e) - (h) depict the energy dependence of
$\chi^{\prime \prime}(\omega$)/$\omega$ calculated by using the fitted
results of
the q-spectra.}
	  \label{f.RvsT-Lt}
	\end{figure}
and integrating over q for the four incommensurate peaks around $(\pi,
\pi)$.  The absolute value of $\chi^{\prime \prime}(\omega)/\omega$ for
each sample
was calculated by using phonon intensities for the same sample as
described in the Appendix.  We present $\chi^{\prime
\prime}(\omega)/\omega$ instead
of $\chi^{\prime \prime}(\omega)$ to reveal any gap structure in the
spectrum because the
latter quantity may go to zero as the energy approaches zero even if
there is no gap. Fig.\@ 4 again exhibits two types of energy
dependencies for the incommensurate peaks for the four samples. There
exists a well-defined energy gap for $x= 0.15$ and 0.18, accompanied
by an enhancement of $\chi^{\prime \prime}(\omega)/\omega$ around 6 meV to
7 meV
while no gap-like structure is seen for $x=0.10$ and 0.25.

The energy dependence of the resolution-corrected q-width of the
incommensurate peak below T$_c$ is shown in Fig. 5.  We
\begin{figure}
	  \centerline{\psfig{file=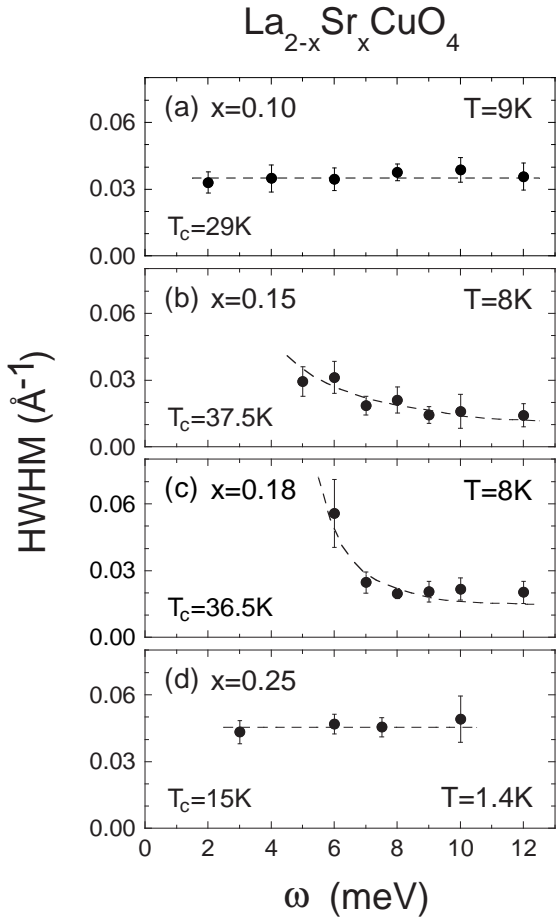,width=\columnwidth}}
	  \caption{Energy dependence of the q-width of an incommensurate peak below
T$_c$ with the instrumental resolution deconvolved.  The q-width in the low
energy region for x=0.15 and 0.18 cannot be obtained due to the opening
of the energy gap below T$_c$.}
	  \label{f.RvsT-Lt}
	\end{figure}
only show the data above 5 and 6 meV for $x=0.15$ and 0.18, respectively
because of the energy gap. We note that the q-width
has a rather different $\omega$ dependence depending on whether or not there is
a gap.  For the gapped samples, above the gap energy the q-width is small and
it increases with decreasing energy as the gap energy is approached.
Furthermore, as will be shown later in Fig. 8, for T $\sim$T$_c$ the q-width
appears to exhibit a peak for excitation energies near the gap energy.  On
the other hand, for
the gapless samples the q-width is constant with energy at least up
to 12 meV.

Next, we show $\chi^{\prime \prime}(\omega)$ for temperatures T$\sim$T$_c$
for $\rm
La_{2-x}Sr_xCuO_4$ crystals with
x=0.15 and 0.18 in Fig. 6.  Although the peak in the intensity appears at
slightly
\begin{figure}
	  \centerline{\psfig{file=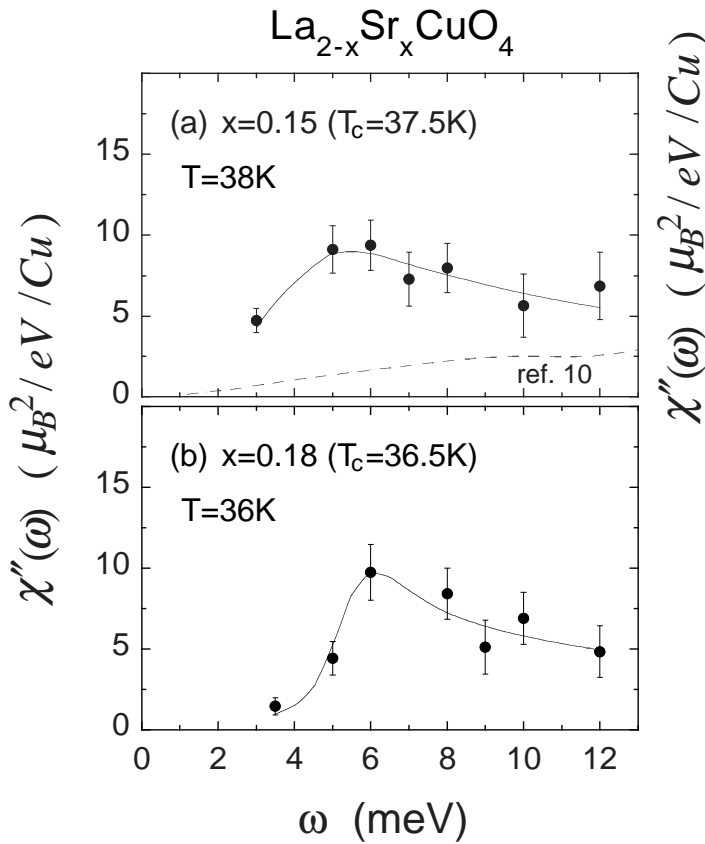,width=\columnwidth}}
	  \caption{Energy dependence of $\chi^{\prime \prime}(\omega)$ for x=0.15(a) and
0.18(b) around T$_c$.  The solid lines are the results of a least squares
fit using
the gap-function, Eq.\@ (4), convoluted with the instrumental resolution
function.  The broken
line in the upper panel is a smoothed interpoltation of the data for
$x=0.14$ reported in ref. [10].}
	  \label{f.RvsT-Lt}
	\end{figure}
lower energies compared to the data below T$_c$, a gap-like structure and
an enhancement in $\chi^{\prime \prime}(\omega)$ still remains.  In the
upper panel
of Fig. 6, we show using a dashed line the data obtained by Hayden
{\it et al.} for the same system with x=0.14 [10].  No peak in
$\chi^{\prime \prime}(\omega)$ is observed in their data.  As shown in Fig.
7, for
\begin{figure}
	  \centerline{\psfig{file=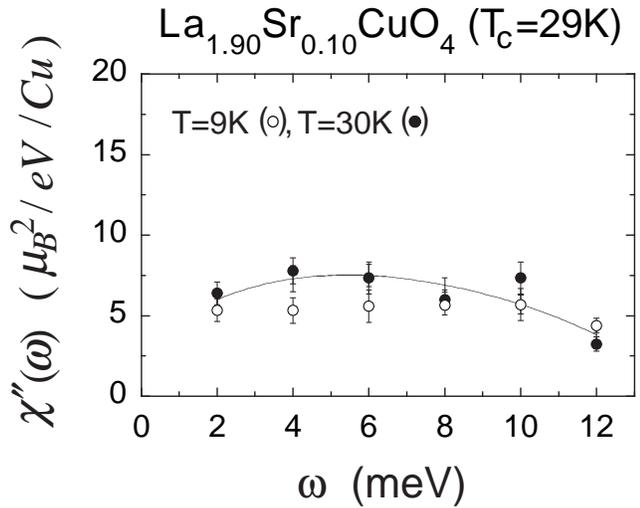,width=\columnwidth}}
	  \caption{Energy spectra of $\chi^{\prime \prime}(\omega)$ for x=0.10 below and
around T$_c$.}
	  \label{f.RvsT-Lt}
	\end{figure}
the gapless sample of $\rm La_{2-x}Sr_xCuO_4$ with x=0.10, both the energy and
temperature variation of $\chi^{\prime \prime}(\omega)$ is weak, although
we see a slight
suppression of $\chi^{\prime \prime}(\omega)$ around 4 to 6 meV at T=9K.

In Fig. 8 we show the energy variation of the q-width of the
incommensurate peak for T$\sim$T$_c$ for x=0.15 and in the inset for
x=0.18.  In the figure, we have plotted results from several different
crystals.  We observe a peak in the q-width at around the same energy
as that at which the enhancement of $\chi^{\prime \prime}(\omega)$ is
observed.  In
\begin{figure}
	  \centerline{\psfig{file=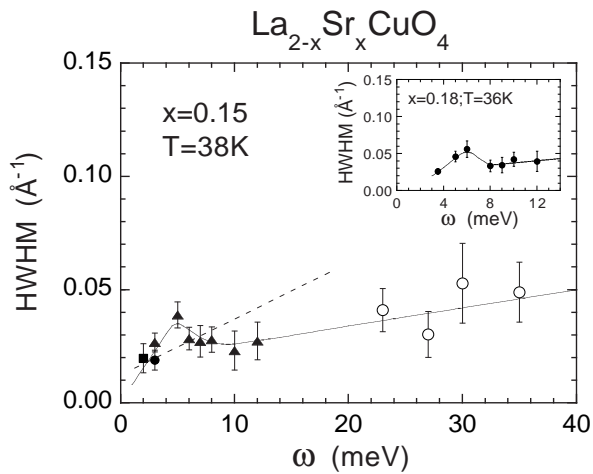,width=\columnwidth}}
	  \caption{Energy dependence of the q-width of an incommensurate peak around
T$_c$ deconvoluted with the instrumental resolution for x=0.15 and
0.18(inset).  The broken line is drawn to represent the data for
x=0.14 reported in ref. [11].}
	  \label{f.RvsT-Lt}
	\end{figure}
contrast, the energy variation of the q-width for the sample with x=0.14
of Aeppli {\it et al.} [11] shows a linear energy dependence.  It should be
noted that the peak width for energies of order 8 to 10 meV is substantially
narrower below
T$_c$ than that at T$_c$ for the gapped samples, particularly for
x=0.18.

\section{Discussion}
\label{sec:level4}
The present study has revealed two types of energy
spectra for the spin fluctuations in Sr-doped superconducting
La$_2$CuO$_4$.  For $x=0.15$ and 0.18, a well-defined
energy gap appears below T$_c$, whereas for the underdoped sample, $x=0.10$, and
highly overdoped sample, x=0.25, no gap is observed even though
they exhibit bulk superconductivity with T$_c$ =29K for $x=0.10$ and
T$_c$=15K for x=0.25.

In order to discuss the results shown in Figure 4 quantitatively, we introduce
a phenomenological dynamical spin susceptibility of the form:

\begin{equation}
\chi^{\prime \prime}(\omega)/\omega =
\left(B{\gamma\over{\gamma^2+\omega^2}}\right)\cdot
Re\left({\omega-i\Gamma}\over{\sqrt{(\omega-i\Gamma)^2-\Delta^2}}\right)
\end{equation}

This corresponds to Lorentzian spin fluctuations in the normal state multiplied
by a gap-function in the superconducting state.  The gap function includes the
gap-energy $\Delta$ together with a broadening $\Gamma$; $\gamma$ is the
inverse
of lifetime of the spin fluctuations.  The parameters B, $\Delta$ and $\Gamma$
are assumed to be constant in the q-region of the incommensurate peaks.  It
should be noted that in Eq.\@ (4), $\chi^{\prime \prime}(\omega)/\omega$ is a simple
Lorentzian in energy for $\Delta=0$.  On the other hand, in the superconducting
state, Eq\@. (4) describes the gap-structure with a non-zero value of
$\Delta$.  As required, the above dynamical spin susceptibility satisfies
detailed balance for the dynamical structure factor S(q, $\omega$).

The observed energy spectra were fitted using Eq.\@ (4) convoluted with the
instrumental energy resolution.  The value of $\Delta$ so-obtained is  $6
\pm 0.3$ meV for both x=0.15 and 0.18 while $\Gamma = 0.6 \pm 0.3$ meV for
x=0.15 and $= 0 \pm 0.2$ meV for x=0.18.  Quite recently, Lake {\it et al.}
reported a similar value for the  energy-gap for a sample of
$La_{2-x}Sr_xCuO_4$
with x=0.16 [25].  We note that $\Delta$ is significantly larger than the value
previously quoted by our collaboration which was around 3.5 meV for x=0.15
[7].  However,
this latter value for the energy-gap was simply the energy below which the
magnetic intensity disappeared entirely below T$_c$.  This energy is
naturally less than our
fitted value simply due to the non-zero broadening of the gap-structure for
x=0.15 together
with the effects of instrumental resolution.  We note that for samples with
x=0.14 and 0.16,
Lake {\it et al.} [25] find
$\Gamma$ = 1.2 meV and 0.1 meV respectively albeit using a somewhat different functional
form than Eq.\@ (4).  Thus, $\Gamma$ decreases
smoothly with
increasing x reaching 0 within the errors for x=0.18 while the gap energy
$\Delta$ appears to
be independent of x for x between 0.14 and 0.18.

In Ref.\@ 6 the gap-energy, $\Delta_m^{inc}$ of the spin fluctuations at the
incommensurate peak position was converted into the value at ($\pi$,0) by
assuming a d(x$^2$-y$^2$)-type q-dependence.  The previous value
$\Delta_m^{inc}$ $\sim$3.5 meV implied a full gap of $\sim$10 meV which is
consistent with the full superconducting gap of $\sim$8 meV inferred from
photoemission spectroscopy [26].  However, the new value of $\Delta_m^{inc}$
determined here implies a full gap of $\sim$18 meV which corresponds to
$\Delta=5.6
kT_c$ which seems too high compared with the BCS weak coupling
value of 1.77 kT$_c$.  This fact, therefore, suggests that
$\Delta_m^{inc}$ may not be simply related to the superconducting gap.  It is
possible that $\Delta_m^{inc}$ instead relates to the so-called pseudo-gap or
the larger energy gaps reported by photoemission and tunneling
spectroscopies.

The gapped samples exhibit an anomalous broadening in the q-width of the
incommensurate peaks for energies around the gap value of $\Delta_m^{inc}$.
Qualitatively, we
can interpret the peak-broadening as arising from the energy independent
dispersion surface of the quasiparticle excitations around the energy-gap.  In
other words, if the magnetic scattering occurs through interband excitations of
quasiparticles, the momentum change can be widely distributed for excitation
energies close to the gap.  As shown in Fig.\@ 8, a broadening in q of the
peak width for
energies of order $\Delta$ is observed at T$_c$ which again indicates the robustness of the 
gap structure of $\chi^{\prime \prime} (\omega)$ for the gapped samples.

We discuss next the possible reasons for the absence of an energy-gap in the gapless
samples with x=0.10 and x=0.25.  At present, it is widely speculated without
any direct experimental evidence that the disorder introduced by dopant
substitution at the La-sites degrades the lifetime of the quasiparticles
and smears out
the gap structure.  However, this is contradicted by the fact that no gap is
observed in either stage-6 or stage-4 $\rm La_2CuO_{4+y}$ in spite of the fact
that structural disorder is minimal in these systems [27,28].  Further, neutron
scattering experiments on the Y1-2-3 system reveal the robustness of the
magnetic
energy-gap against Zn-impurities [29,30].  In the Zn-doped Y1-2-3 system,
although low energy
spin fluctuations appear at low temperatures, the gap-like or pseudo-gap
structure
appears at higher temperatures.

For the x=0.10 sample, the low energy spin excitations are simply explained as
originating from the spin wave excitations concomitant with the elastic
magnetic order which occurs in the superconducting state.  Such order is
observed for all x between 0.02 and 0.135 in $\rm La_2CuO_4$.  However, a
different
explanation is required for the highly overdoped gapless sample, x=0.25,
because in that case
there is no evidence for magnetic order coexisting with the
superconductivity.  Here, we
simply point out that there is a dramatic degradation of the spacial
coherence of the spin
correlations in the overdoped region particularly near the upper critical
doping value for the
superconductivity as pointed out by Yamada et al. [15].  In this doping region, the spatial coherence
length for
the low energy spin fluctuations becomes comparable to or shorter than the
superconducting coherence length.  Therefore, the energy gap in this region is
probably vitiated by the degradation of the lifetime of quasiparticles due to
the short ranged spin fluctuations.

\section{Summary}
\label{sec:level5}
We have studied the energy dependence of the spin
fluctuations for superconducting La$_{2-x}$Sr$_x$CuO$_4$ ($x$=0.10,
0.15, 0.18, 0.25) using neutron inelastic scattering.  Below T$_c$ a
well-defined energy gap in the incommensurate spin fluctuations is
observed in the superconducting state of the optimally doped
($x=0.15$) and slightly over doped ($x=0.18$) samples.  An
enhancement in $\chi^{\prime \prime}(\omega)$ caused by the peak-broadening
is also
observed at $\sim$6 meV which remains even at T$_c$.  For the
underdoped sample with $x=0.10$ and the highly overdoped sample with $x=0.25$,
no clear gap is ovserved even though these samples show bulk superconductivity.

\begin{center}
{\bf Acknowledgements}
\end{center}

The authors acknowledge K. Nemoto and M.
Onodera for their technical assistance at JAERI and Tohoku
University.  We wish to thank H. Fukuyama, T. Tanamoto, H. Kohno, S.
Hosoya, K. Hirota and H. Kimura for valuable
discussions.  We also thank S. Wakimoto for his help in crystal growth of 
x=0.15 samples.  The present work was supported in part by a Grant-In-Aid
for Scientific Research from the Ministry of Education, Science,
Culture and Sports of Japan and a Grant for the Promotion of Science
from the Science and Technology Agency of CREST.  Work at Brookhaven
National Laboratory was carried out under contract No.
DE-AC02-98CH10886, Division of Material Science, U. S. Department of
Energy.  The research at Massachusetts Institute of Technology was
supported by the National Science Foundation under Grant No.
DMR97-04532 and the MRSEC Program of the National Science Foundation under
Award No. DMR98-08941.

\begin{center}
{\bf Appendix}
\end{center}

We have converted the observed magnetic intensity into
the absolute value of $\chi^{\prime \prime}(\omega)$ by comparing with the
intensity
of phonon scattering.  The phonon intensity is given by
\begin{equation}
I_\nu
=\alpha\cdot\left({d\sigma}\over{d\Omega}\right)
\end{equation}
\noindent  where
$\alpha$ is the detector efficiency and $I_\nu$ is the
energy-integrated phonon intensity measured by constant-Q scans.  The
energy integrated phonon scattering cross section is given by:
\begin{equation}
\left({d\sigma}\over{d\Omega}\right)=C{{|F|^2}\over{M}}{{|Q|^2 cos^2
\beta\cdot N}\over{\hbar\omega}}e^{-2W}<n+1>
\end{equation}
\noindent where C=2.09
barns, $F$ is the nuclear structure factor, $M$ is the total atomic
mass in a unit cell, $\beta$ is the angle between Q and the phonon
polarization direction, $N$ is the number of unit cells, $e^{-2W}$ is the
Debye-Waller factor and $<n+1>$ is the population factor.  We have
measured the integrated inensity of the acoustic phonons by constant-Q
scans at (2, -0.13, 0) for $x=0.10$ and (2, -0.17, 0) for $x=0.15$,
0.18, and 0.25.  The magnetic scattering intensity is measured by
constant-$\omega$ scans.   To scale the magnetic scattering intensity
with the phonon intensity measured in a different scan mode, $I_\nu$ is
divided by the sound velocity.  The absolute value of
$\chi^{\prime \prime} (q,\omega)$ is estimated using the following
magenetic cross
section per formula unit for a Heisenberg system without N\'eel
ordering:
\begin{equation}
\left({d^2\sigma}\over{d\Omega
d\omega}\right)=r^2_0{{k_f}\over{k_i}}\left[{g\over2}f(Q)e^{-W}\right]^2
\sum_\alpha (1-\hat k^2_\alpha)S^\alpha(Q,\omega)
\end{equation}

\begin{equation}
S^\alpha(Q,\omega)={1\over\pi}{1\over{g^2\mu^2_B}}<n+1>Im
\chi^\alpha(Q,\omega)
\end{equation}
Here $r^2_0=0.291$ barns, $f(Q)$ is magnetic form factor and $\hat k =
Q/|Q|$.

\end{document}